\documentclass[11pt,a4paper]{article}
\usepackage{jcappub}

\usepackage{epsfig,psfrag,graphics,verbatim}
\usepackage{graphicx}

\usepackage{dcolumn}
\usepackage{bm}
\usepackage{multirow}
\usepackage{epsfig}
\usepackage{amsmath,amssymb}
\usepackage{psfig}
\usepackage{ulem}
\usepackage{natbib}

\def \apj  {ApJ}

\def \mnras {MNRAS}
\def \chisq  {\ifmmode  \chi^2   \else  $\chi^2$  \fi}  
\def \spose#1{\hbox  to 0pt{#1\hss}}  
\def \lta{\mathrel{\spose{\lower 3pt\hbox{$\sim$}}\raise  2.0pt\hbox{$<$}}}
\def \gta{\mathrel{\spose{\lower  3pt\hbox{$\sim$}}\raise 2.0pt\hbox{$>$}}}

\title{Density Profiles of CDM Microhalos and their Implications for Annihilation Boost Factors}

\author{Donnino Anderhalden \&}
\author{Juerg Diemand}

\affiliation{Institute for Theoretical Physics, University of Z\"urich, Winterthurerstrasse 190, 8057 Z\"urich, Switzerland}
\emailAdd{donninoa@physik.uzh.ch}
\emailAdd{diemand@physik.uzh.ch}

\abstract{In a standard cold dark matter (CDM) cosmology, microhalos at the CDM cutoff scale are the first and smallest objects expected to form in the universe. Here we present results of high resolution simulations of three representative roughly Earth-mass microhalos in order to determine their inner density profile. We find that CDM microhalos in simulations without a cutoff in the power spectrum roughly follow the NFW density profile, just like the much larger CDM halos on galaxy and galaxy cluster scales. But having a cutoff in the initial power spectrum at a typical neutralino free streaming scale of $10^{-7} M_{\odot}$ makes their inner density profiles considerably steeper, i.e. $\rho \propto r^{-(1.3-1.4)}$, in good agreement with the results from Ishiyama et al. (2010). An extrapolation of the halo and subhalo mass functions down to the cutoff scale indicates that microhalos are extremely abundant throughout the present day dark matter distribution and might contribute significantly to indirect dark matter detection signals. Assuming a transition from a NFW to a steeper inner profile ($\rho \propto r^{-1.4}$) two orders of magnitude above the cutoff scale, the total boost factor for a Milky Way sized dark matter halo increases from about 3.5 to 4. We further find that CDM microhalo concentrations are consistent with the Bullock et al. (2001) model and clearly rule out simplistic power law models for the mass dependence of concentrations and subhalo annihilation, which would erroneously lead to very large boost factors (a few hundred for galaxy halos and over 1000 for clusters).}


\begin{document}
\maketitle

\section{Introduction}
The standard model of cosmology is characterised by a hierarchical bottom-up formation of structure \citep{Peebles1982}, in which the size of the smallest objects is set by the free streaming length of the dark matter particle. For a standard supersymmetric cold dark matter (CDM) candidate with a mass of 100 - 1000 GeV, free streaming leads to a cutoff in the matter power spectrum at a scale in a range of about $10^{-12}$ to $10^{-3}$ M$_{\odot}$ \citep[e.g.][]{KolbTkachev96,Hofmann2001,Green2004,Bertone2005,LoebZaldarriaga2005,Bertschinger2006,Profumo2006,Bringmann2009}. Such microhalos, and especially their inner regions,  might be dense enough  to survive tidal disruption up to present time and they are therefore potential contributors to indirect dark matter detection signals \cite[][]{Berezinsky2003,Berezinsky2006,Diemand2005,Goerdt2007,Zhao2007,GreenGoodwin2007,Schneider2010,Koushiappas2006,Koushiappas2009,Kamionkowskietal2010}.

The density profiles of CDM halos from dwarf galaxy to galaxy cluster scales have been studied extensively with many high resolution cosmological simulations \citep[e.g.][]{Navarro2004,Diemand2004profiles,Diemandetal2008,Springeletal2008,stadel09}. All CDM halos on theses scales roughly follow the universal NFW profile \cite{NFW}. The NFW function has only one free parameter, the scale radius $r_s$ or concentration $c = r_{\rm virial}/r_s$, which is directly related to the initial mass fluctuation spectrum and the typical halo formation times \cite{Bullock2001,Kuhlen2005,Maccio2008,Pradaetal2012}.
Unlike these well studied massive CDM halos, the first, smallest microhalos do not form through hierarchical merging and they do not contain any subhalos. Their internal structure might be different from that of cluster- or galaxy-sized dark matter halos.
The first numerical simulations of microhalo formation \cite{Diemand2005} did not have the resolution to resolve the central regions of microhalos, and in the well resolved radial range the microhalos showed similar profiles as those found in large CDM halos. Recently, Ishiyama et al. simulated microhalo formation with 20 times better mass resolution and surprisingly steep inner slopes were found for halos near the cutoff scale \cite{Ishiyama2010}. Simulated without the cutoff in the initial power spectrum, the same microhalos turned out to have shallower NFW-like inner profiles. In this work we simulate the formation of microhalos with a mass resolutions as low as $m_p = 1.46\times10^{-14}$ M$_{\odot}$, 64 times higher than Ishiyama et al. \cite{Ishiyama2010}, and we are able to largely reproduce their surprising results about the steeper density profiles near the cutoff scale.
Substructure in the form of microhalos contributes significantly to the dark matter annihilation signals from all larger scale dark matter objects (halos, subhalos, galactic center). Here we show that the steeper mircohalo density profiles might increase the substructure boost factor of a galaxy sized CDM halo by about 5 to 12 per cent.

\section{$N$-body Simulations}\label{nbodysims}
Numerical simulations have been performed with the parallel treecode {\sc pkdgrav}, written by Joachim Stadel and Thomas Quinn \citep[][]{Stadel2001}. Initial conditions are generated with a parallel version of the GRAFIC package \citep[][]{Bertschinger2001} and are based on the cosmological parameters taken from WMAP7 \citep[][]{Komatsuetal2011}: $\sigma_8=0.8$, $h=0.7$, $\Omega_{\rm dm}=0.227$, $\Omega_{\rm b}=0.046$, $\Omega_{\Lambda}=0.727$ and $n_{\rm s}=0.961$. Our simulations start at an initial redshift of $z_{\rm ic} = 500$ and run until $z_{\rm f} = 31$, have a force softening length $\epsilon$ of $1/50$ times the mean inter-particle separation and a fixed time-stepping with $N_{\rm steps}=10`000$.

To begin with, we have run two cosmological boxes of 30 pc and a total particle number of $N=512^3$, both having identical random seeds. One of the simulations contains an exponential cutoff in the matter power spectrum corresponding to a CDM particle mass of $\sim 1$ TeV \citep[][]{Green2004}. We use the friends-of-friends algorithm \citep[][]{Stillinger1963,Davis1985} with a linking lenght of $b = 0.2$ to identify the halos at $z_{\rm f} = 31$. Then we select the three largest halos and rerun them at 8 times better mass resolution, again with and without cutoff.  These refined simulations have a mass resolution of $m_p = 9.3 \times10^{-13}$ M$_{\odot}$ and a softening length of $\epsilon \simeq 5\times10^{-5}$ pc, which is the same as in the Ishiyama et al. (2010) simulations \cite{Ishiyama2010}.

In order to test the very inner structure of microhalos right at the cutoff scale, we rerun one of the three halos once more with further increased resolution, this time only for the case with the cutoff in the power spectrum. This ultrahigh resolution microhalo contains approximately $33\times 10^6$ particles within $r_{200}$ with a simulated particle mass of $m_p = 1.46\times10^{-14}$ M$_{\odot}$ (64 times better mass resolution than in \cite{Ishiyama2010}) and a softening length of $\epsilon \simeq 1.25\times 10^{-5}$ pc. For what follows, we will name the two levels of resolution {\it Level 1} and {\it Level 2} respectively. To check for numerical convergence we did rerun our {\it Level 1} simulation with different starting redshift ($z_{\rm ic} = 1`000$), force resolution ($\epsilon/2$) and time-stepping ($N_{\rm steps}=5`000$), see Section \ref{sec:profiles} and the last panel in Figure \ref{densityprofileslowres}.
\begin{figure}
\begin{center}
\includegraphics[scale=0.1338]{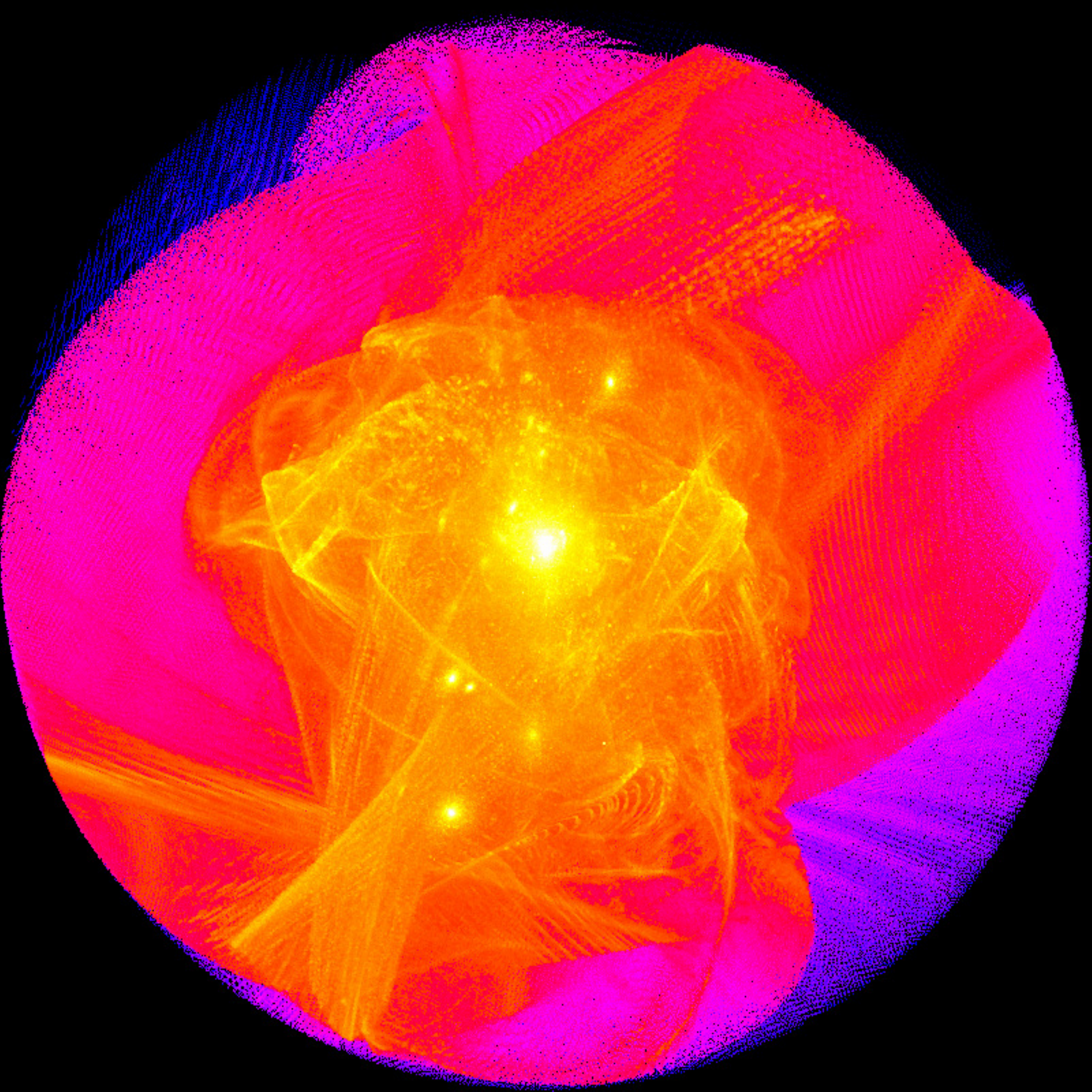}
\includegraphics[scale=0.1338]{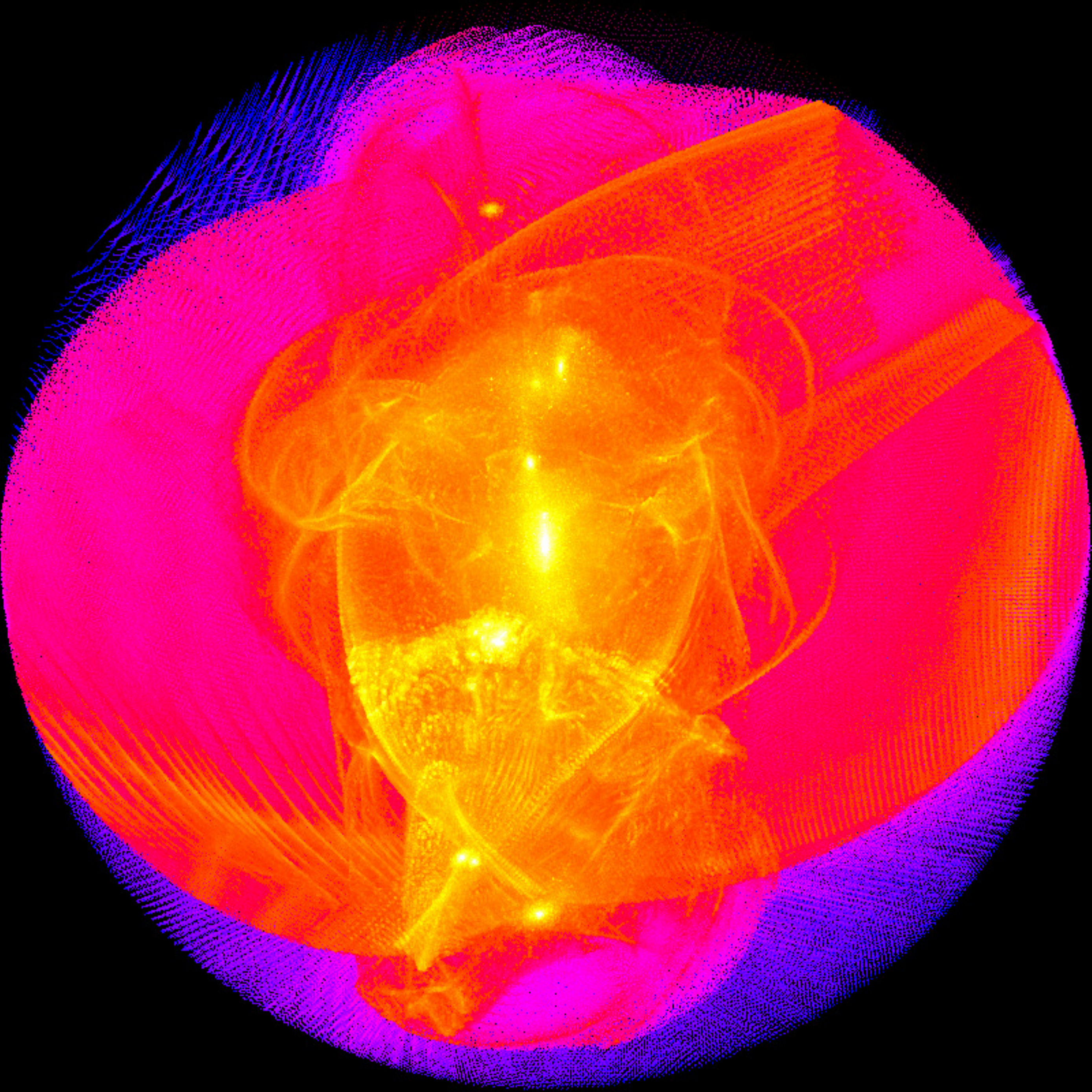}
\includegraphics[scale=0.1338]{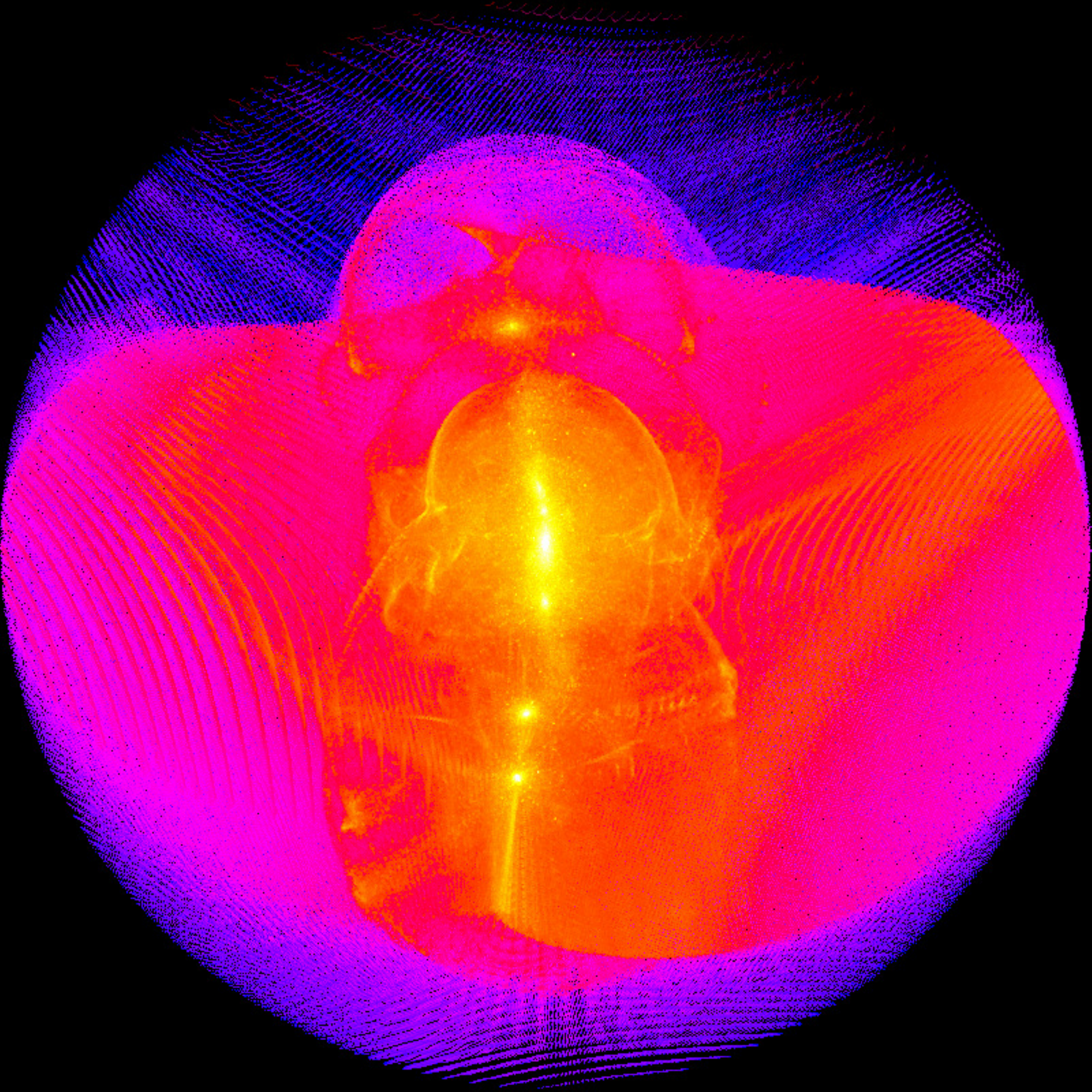}
\includegraphics[scale=0.1338]{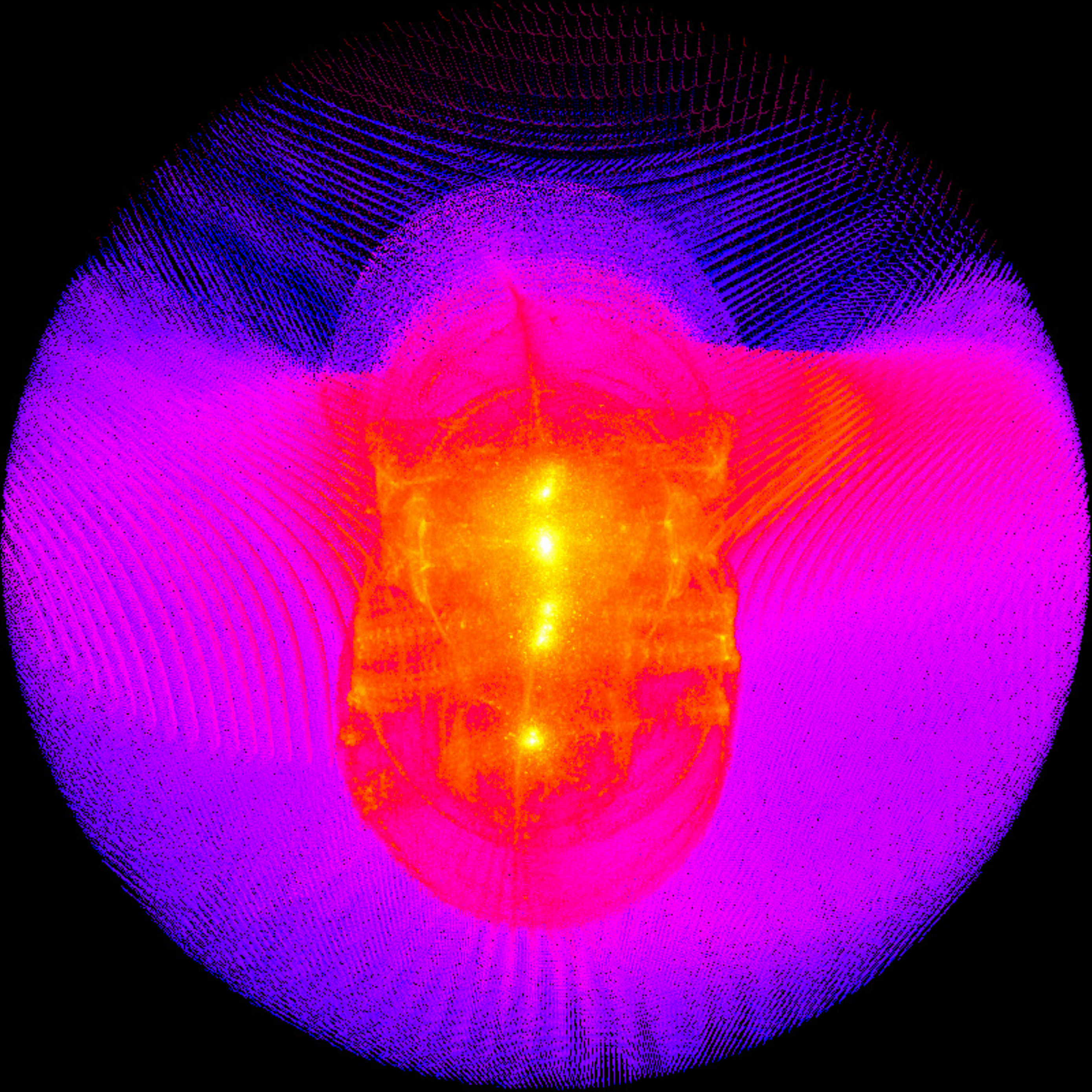}\\
\includegraphics[scale=0.1338]{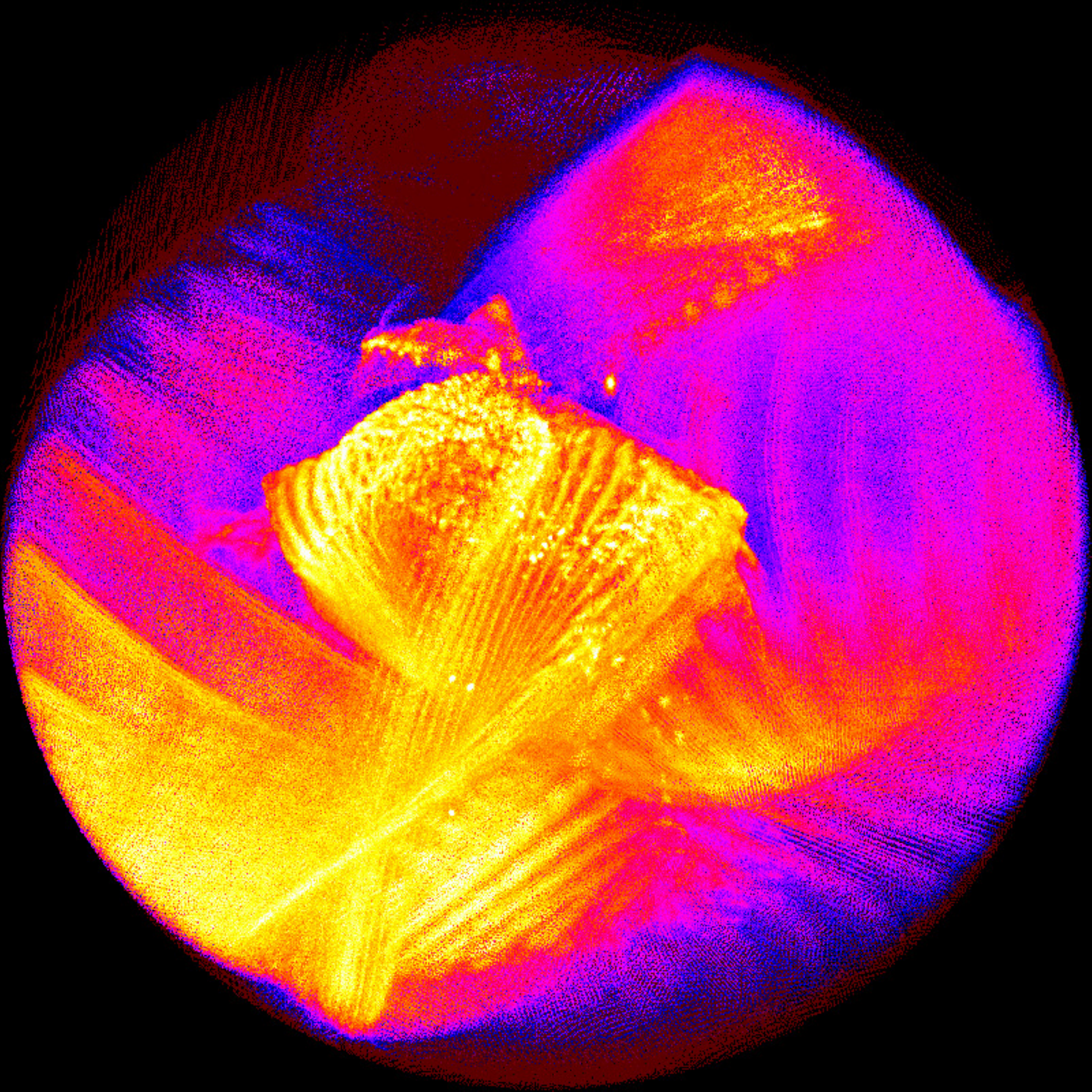}
\includegraphics[scale=0.1338]{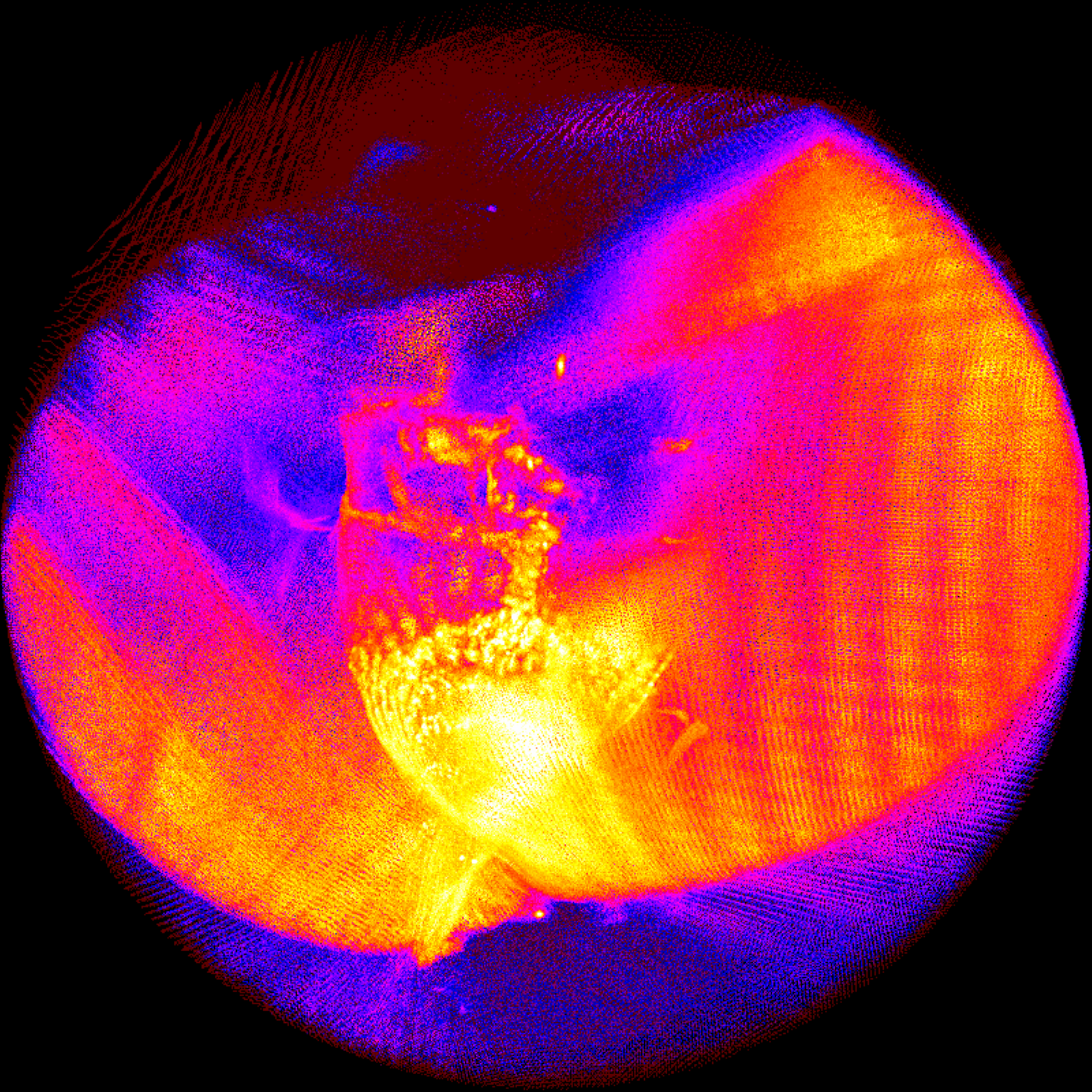}
\includegraphics[scale=0.1338]{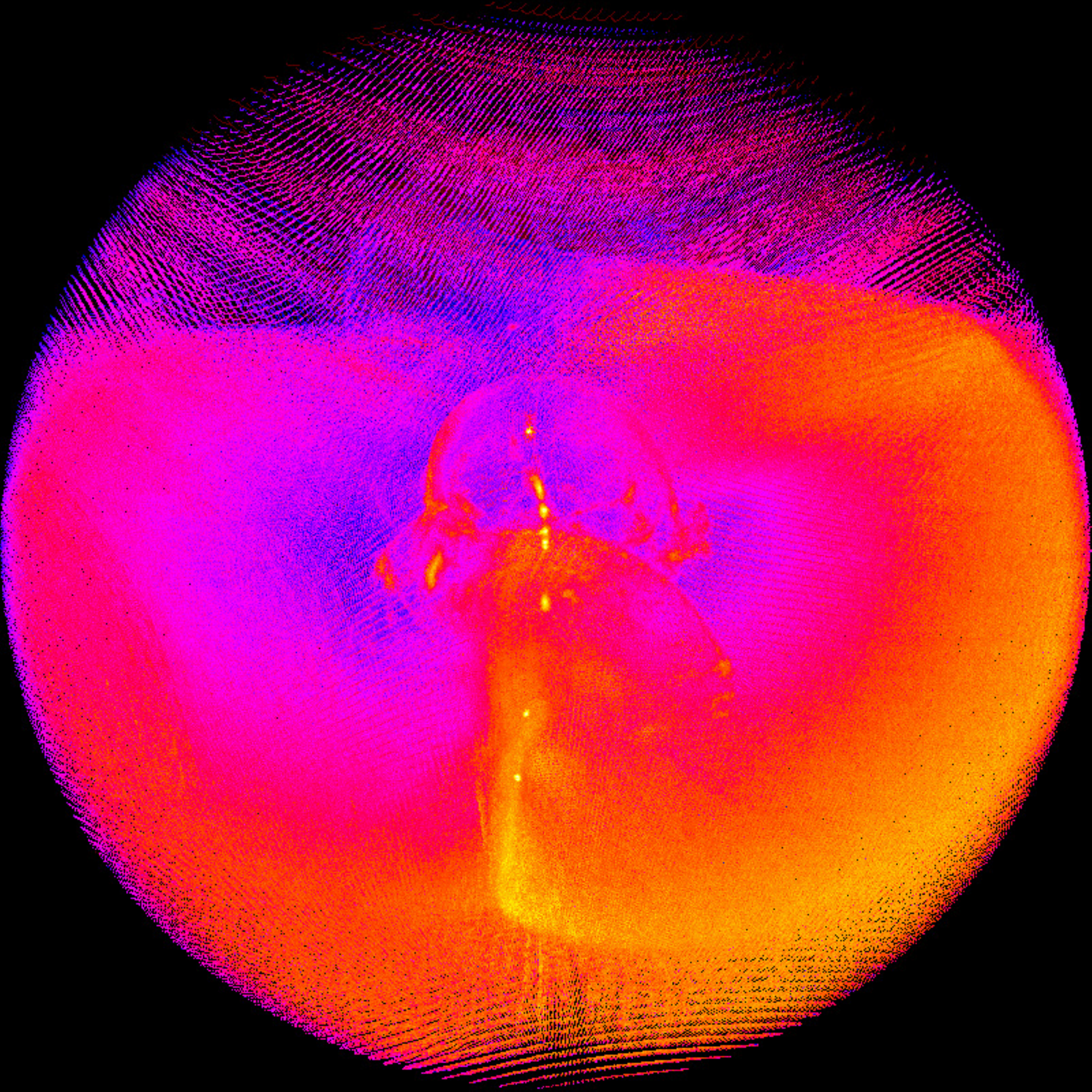}
\includegraphics[scale=0.1338]{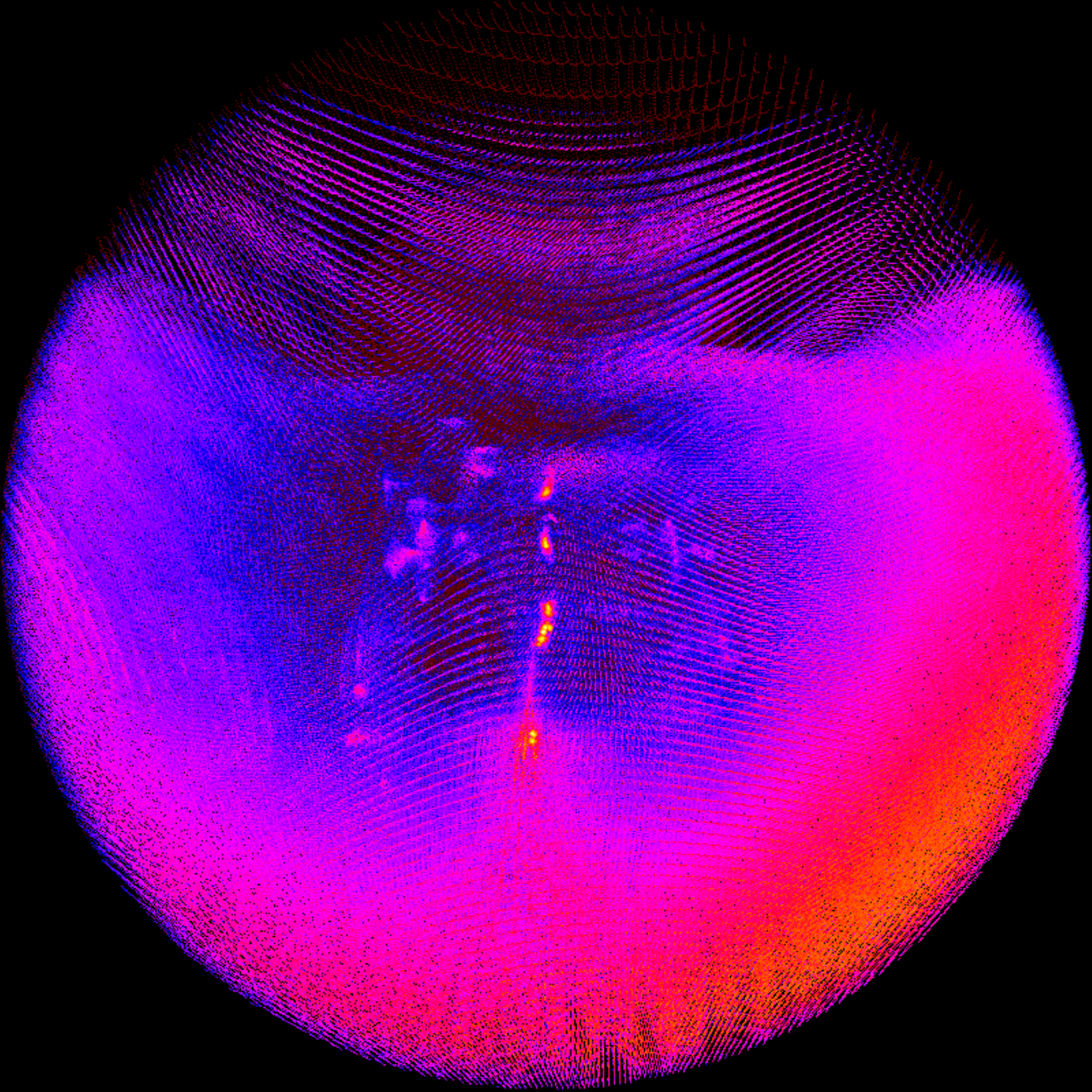}
\caption{Top: Projected density maps of the high resolution microhalo, at redshifts (from left to right): $z=31$, $z=33$, $z=39$, $z=43$.  Bottom: corresponding phase space density maps, calculated with Enbid. In all panels, (phase space) densities are represented in a logarithmic color map: bright regions refer to a high (phase space) density whereas dark regions refer to a low (phase space) density.}
\label{densitymaps}
\end{center}
\end{figure}

Figure \ref{densitymaps} shows the projected density (top-row) and phase-space (bottom-row) density maps of the {\it Level 2} halo at four different redshifts ($z=31,\; 33,\; 39,\; 43$). At $z=31$, there are almost no subhalos, but there is abundant substructures in the form of caustics, similar to those in the classical self-similar secondary infall model \citep[][]{FillmoreGoldreich,Bertschinger1985}. Such caustics could increase the dark matter annihilation rate by contributing to the total annihilation luminosity (see Section \ref{sec:boost}) \cite{Mohayaee2006,Mohayaee2007,DiemandKuhlen2008}. To estimate this contribution for the present {\it Level 2} halo, we compare the signal coming from each individual simulation particle within the halo ($L_1 \sim \sum_i \rho_i m_i$), with the one coming from when binning up the particles in spherical shells and integrating over the volume ($L_2 \sim \int \rho(r)^2_{\rm bin} dV$). We find the ratio $L_1 / L_2$ to be 3.5 at $z=43$, and decreasing to roughly 2.3 at $z=31$. By smoothing out these caustics (i.e. increasing the number of considered neighbours from 32 to 1`000 when computing the local densities), $L_1 / L_2$ shrinks to 1.5 at $z=31$. This suggest that at $z=31$ the microhalo annihilation luminosity is increased by a factor of about $1.5 \simeq 2.3/1.5$ due to caustics and by another factor of 1.5 because of other departures from spherical symmetry. It is possible that the actual contribution from caustics is significantly larger than that, as we only start to resolve caustics in our highest resolution simulations and more detailed future simulations might resolve more and sharper caustics. However, we do see that relative contribution of caustics to the microhalo annihilation signal decreases with time and it is difficult to estimate how large it will be at $z=0$. Therefore we decided to ignore their potential contribution when calculating the boost factors at z=0 in Section \ref{sec:boost}.

\begin{figure}
\begin{center}
\includegraphics[scale=0.37]{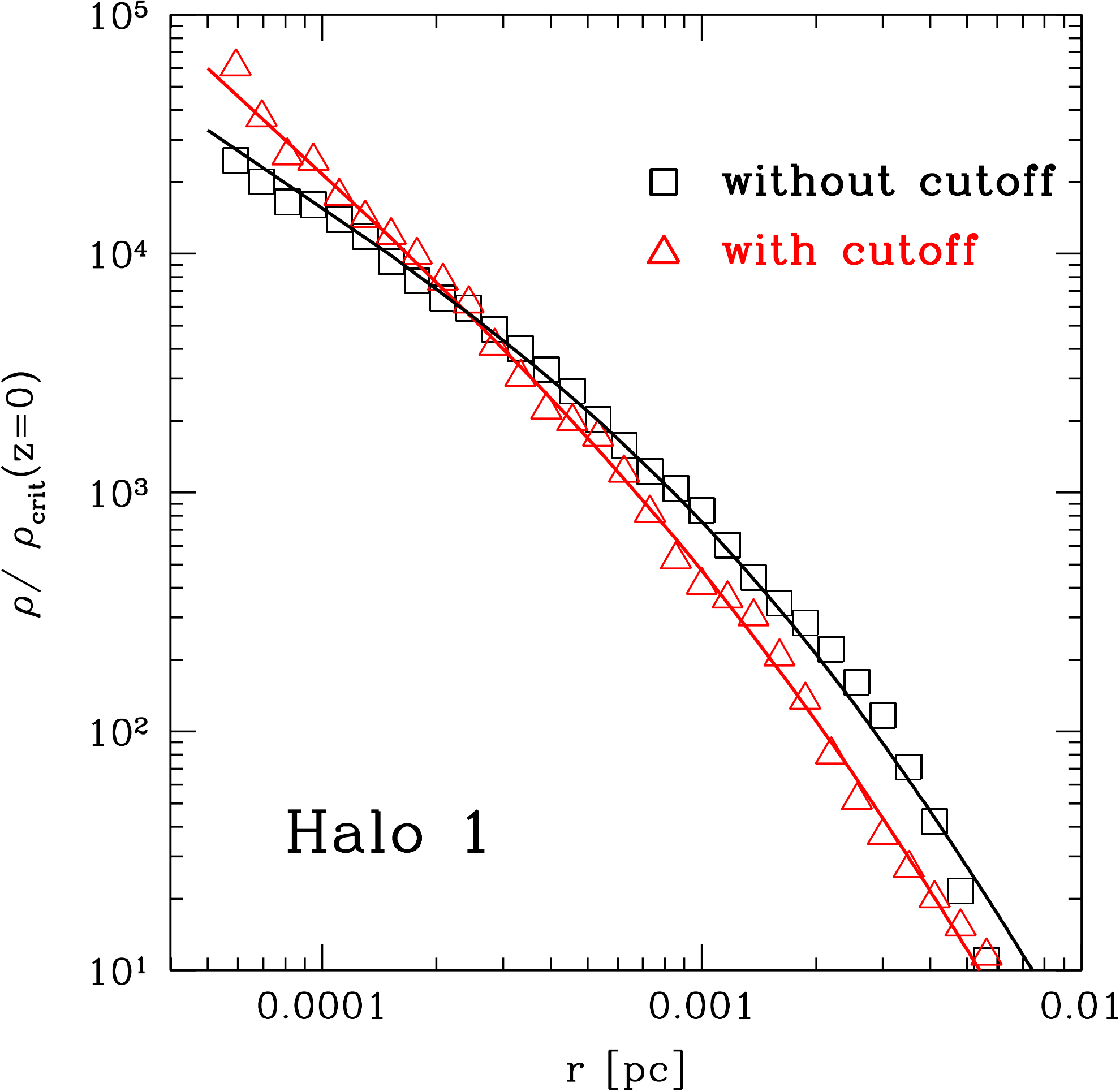}
\hspace{2mm}
\includegraphics[scale=0.37]{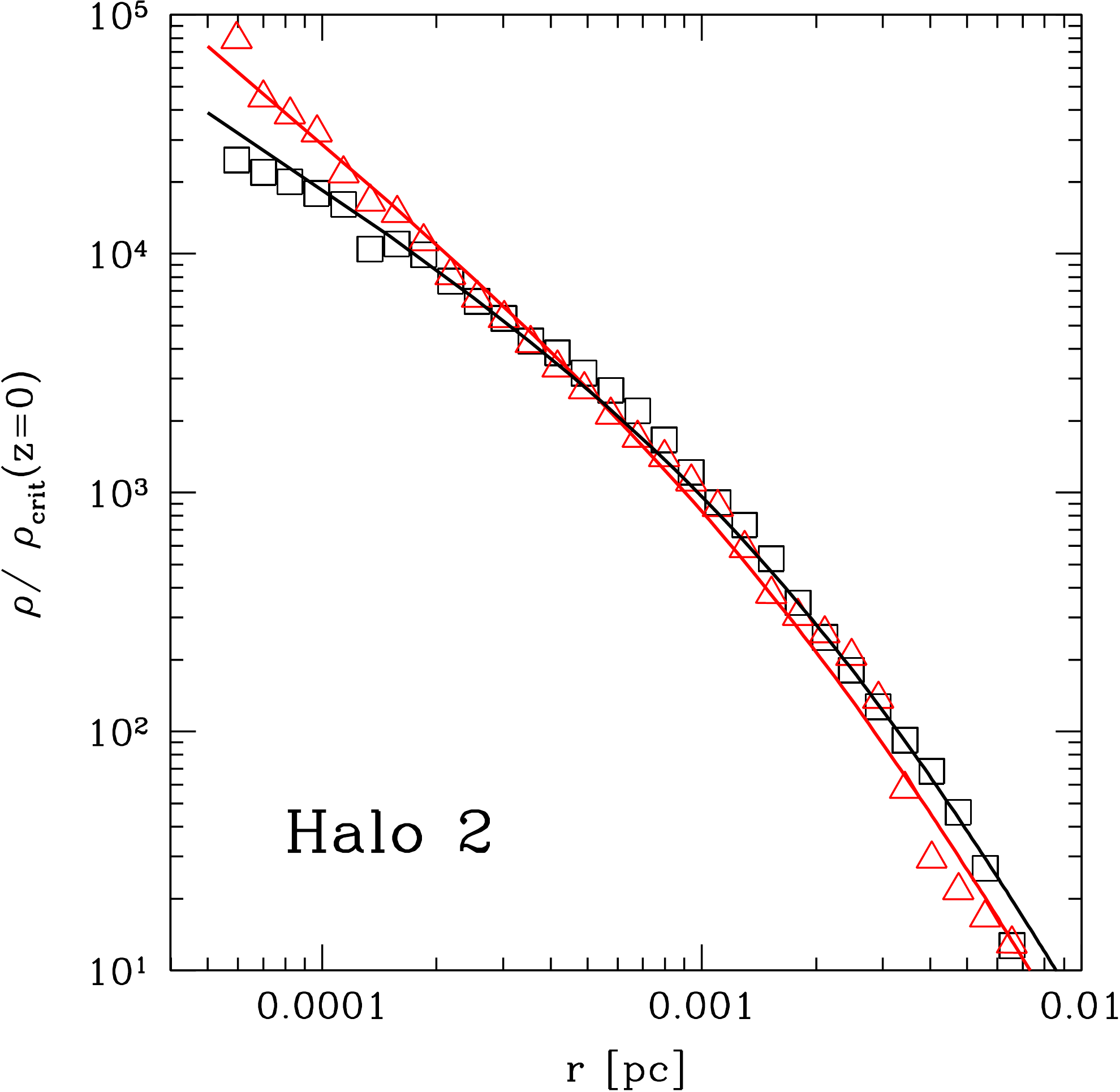}\\
\includegraphics[scale=0.37]{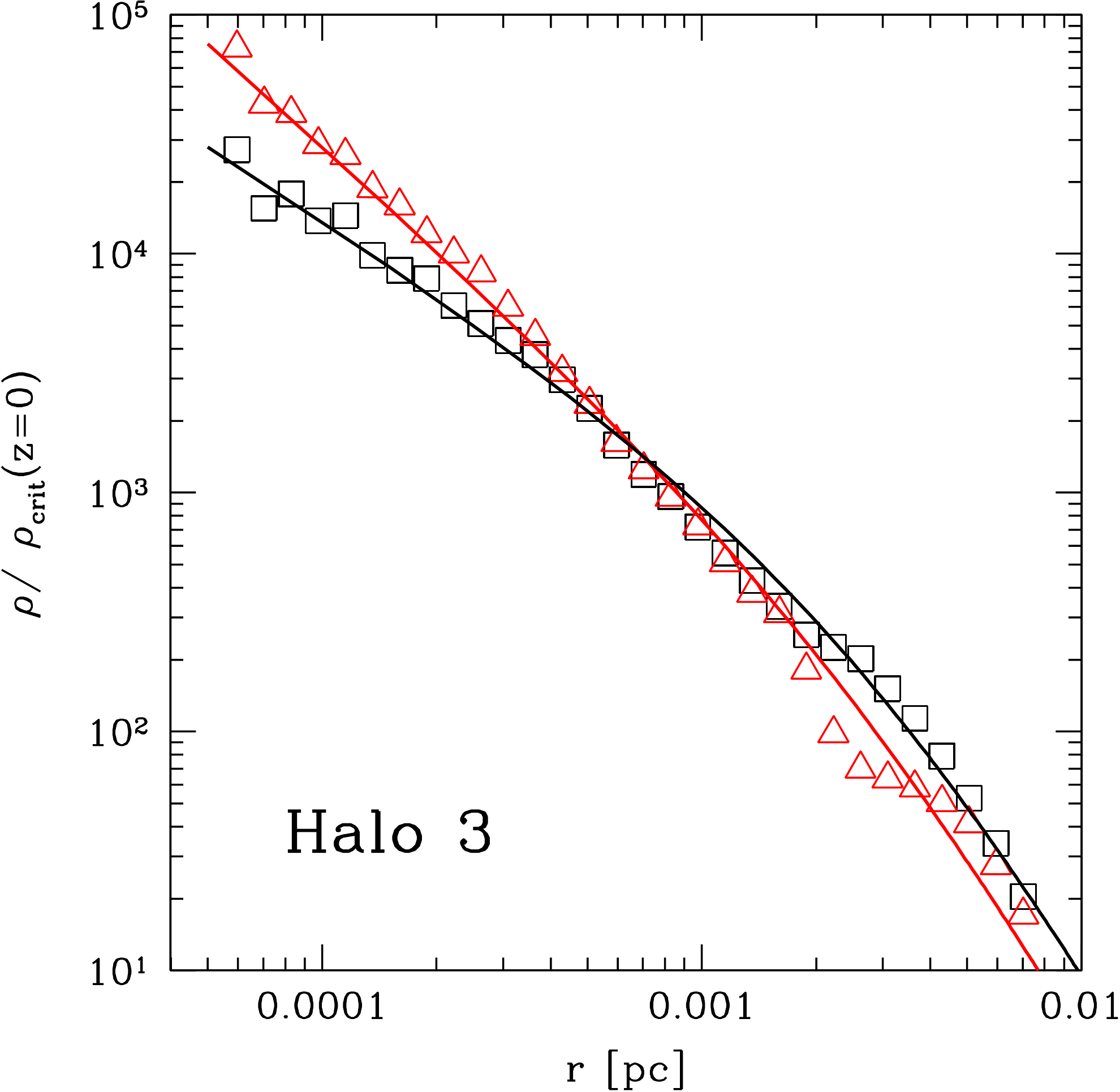}
\hspace{2.3mm}
\includegraphics[scale=0.37]{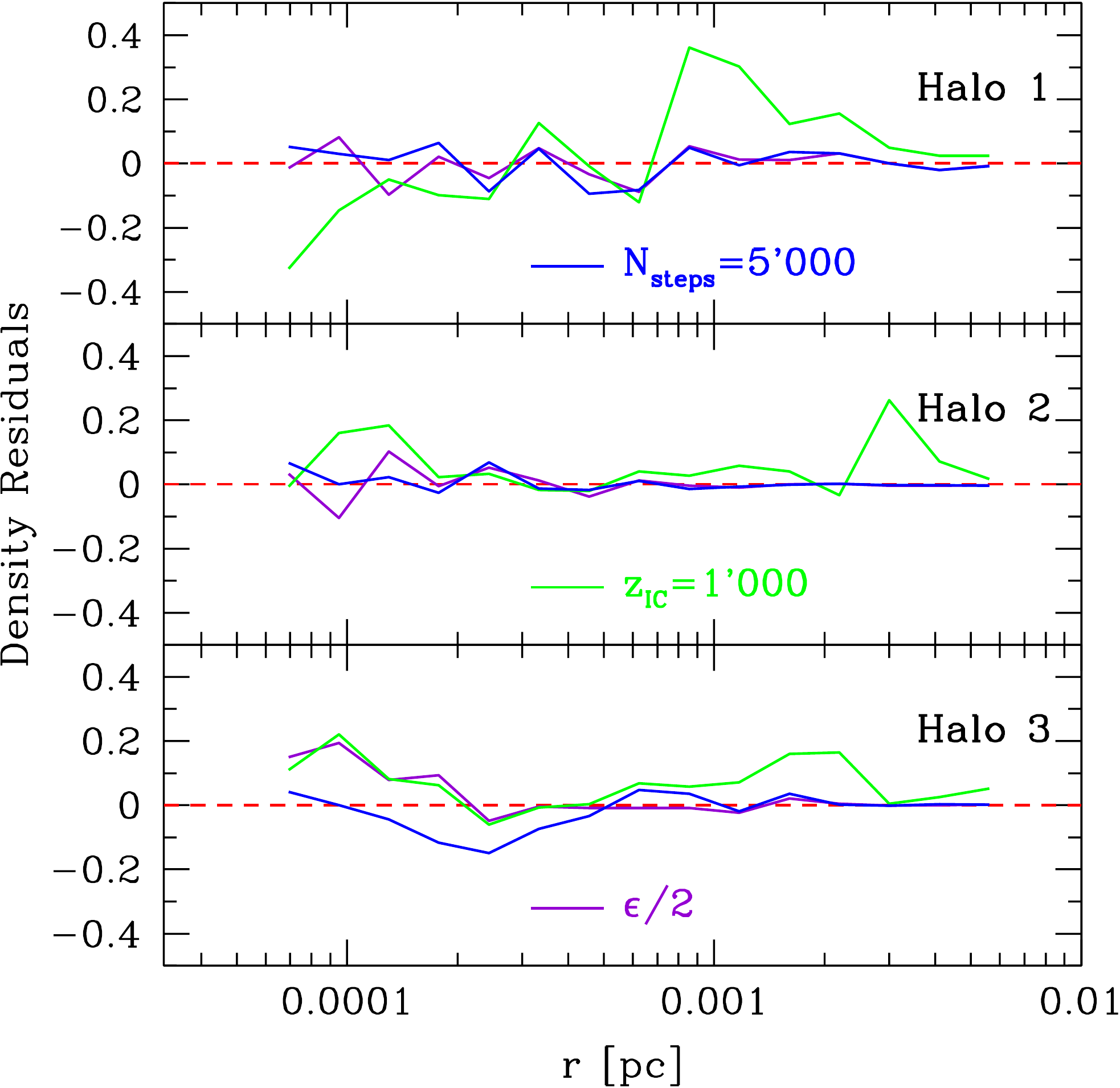}
\caption{Panels 1-3: Spherically averaged density profiles of the three largest collapsed microhalos at $z=31$, with (red triangles) and without (black squares) cutoff. The red solid lines refer to the best fit according to Eq. \eqref{densityfit} with $\alpha = 1.4$ (Halo 1 \& Halo 3) and $\alpha = 1.3$ (Halo 2), the black solid lines refer to a NFW fit respectively. The radial distance is plotted in physical units, densities in units of $\rho_{\rm crit}$ at $z=0$. Panel 4: Density residuals between the {\it Level 1} run and three convergence test simulations, each varying one simulation parameter.}
\label{densityprofileslowres}
\end{center}
\end{figure}
\section{Microhalo density profiles}\label{sec:profiles}

In order to quantify the detectability of microhalos (e.g. \cite{Ishiyama2010}) and the survival probability as galactic subhalos \cite[][]{Berezinsky2003,Berezinsky2006,Diemand2005,Goerdt2007,Zhao2007,GreenGoodwin2007,Schneider2010}, one needs to know the form of their density profiles and their typical concentrations (we use the concentration definition $c_{200} = r_{200} / r_s$, where $r_{200}$ is the radius enclosing 200 times the critical density of the Universe).

\subsection{The inner density profiles}

Figure \ref{densityprofileslowres} shows the spherically averaged density profiles of the three largest {\it Level 1} microhalos (see Table \ref{halosizetable} for details) with (red triangles) and without (black squares) cutoff. It is evident that the inner slope of the models with cutoff is substantially steeper than without the cutoff. In order to quantify the central structure of the three halos in the cutoff model, we have fitted the density profiles with the following conventional parametrisation \citep[e.g.][]{NFW}:
\begin{equation}\label{densityfit}
\rho(r) = \frac{\rho_s}{\Big(\frac{r}{r_s}\Big)^\alpha \Big(1 + \frac{r}{r_s}\Big)^{3 - \alpha}}.
\end{equation}
Using the Levenberg \& Marquardt method, we obtained the following values for the slope of the inner profile via a $\chi^2$ minimisation:
\begin{equation}
\alpha_{\rm [Halo\;1]} \simeq 1.4, \quad \alpha_{\rm [Halo\; 2]} \simeq 1.3, \quad \alpha_{\rm [Halo\;3]} \simeq 1.4.
\end{equation}
The fits are shown by the red solid lines in Fig. \ref{densityprofileslowres} (panels 1-3). On the other hand, a NFW profile ($\alpha = 1$) provides a reasonable fit for the microhalo densities in the model without a cutoff (black solid lines).

In order to exclude a systematic computational error leading to such steep profiles, we have rerun the {\it Level 1} simulation in three variations to verify numerical convergence: (i) with half as much time-steps ($N_{\rm steps}=5000$), (ii) with a starting redshift of $z_{\rm ic}=1`000$ and (iii) with a reduced gravitational softening length ($\epsilon/2$). The resulting density profiles are represented as residuals ($\rho_i/\rho - 1$) in the last panel of Figure \ref{densityprofileslowres}. In all three halos, the deviations are negligible for the $N_{\rm steps}=5`000$ (blue solid lines) and the $\epsilon/2$ (green solid lines) simulation. The run with increased starting redshift (red solid lines) shows the largest fluctuations, most probably due to the fact that force errors become significant in the nearly uniform, very high redshift matter density field. However, even for the latter case, differences are only of the order of 20 per cent. We conclude that our results are numerically robust, and that the origin of such steep density cusps must indeed be physical. The different inner profiles of microhalos near the cutoff scale is most likely related to their different, non-hierarchical formation, but we are currently unable to explain this surprising result.

\begin{figure}
\begin{center}
\includegraphics[scale=0.5]{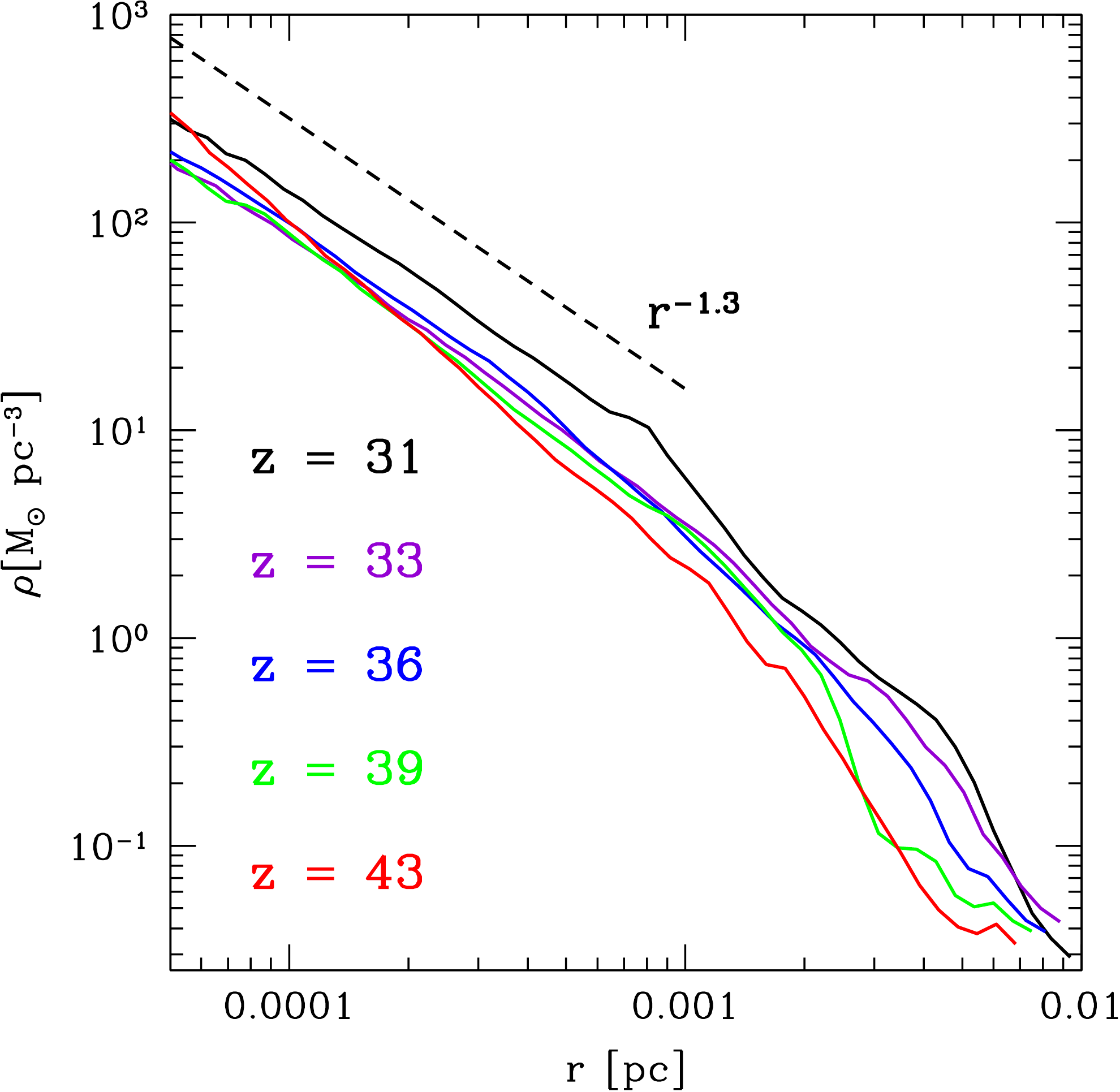}
\caption{Redshift evolution of the spherically averaged density profile in the {\it Level 2} refinement of Halo 2. The black dashed line indicates the inner slope of $\alpha = 1.3$. Distance and density are plotted in physical units.}
\label{densityevolution}
\end{center}
\end{figure}

Halo 2 was resimulated at the {\it Level 2} refinement, i.e. with 64 times better mass resolution. This allows us to resolve the very inner part of the density profile, down to a resolved radius of $\sim 5\times 10^{-5}$ pc. Even at these very small scales we do not observe any flattening or deviation from a $\alpha=1.3$ cusp, as indicated by the dashed black line in Figure \ref{densityevolution}. Further the evolution of the density profile from $z=43$ to $z=31$ is shown. The inner slope of the microhalo remains constant after about $z=39$.

\subsection{Microhalo concentrations}\label{concentrations}
\begin{table}
\caption{Halo parameters of the {\it Level 1} simulation at redshift $z=31$. $M_{200}$ and $r_{200}$ are measured as 200 times the critical density, $\alpha$ is the inner density slope of the measured density profile (see Eq. \eqref{densityfit}), $\alpha = 1$ corresponds to the NFW profile. Distances are given in physical units.}
\label{halosizetable}
\begin{center}
\begin{tabular}{lc|cccccc}
\hline
\hline 
 &  & $M_{200}$ & $r_{200}$ & $r_s$ & $c_{200}= r_{200}/r_s$ & $c_{\rm NFW}$ &$\alpha$  \\ 
&  & [$10^{-7}$ M$_{\odot}$] & [$10^{-3}$ pc] &$[10^{-3}$ pc]  & & &\\
\hline\hline
\multirow{3}{*}{Cutoff} & Halo 1 & 1.22 & 5.63 & 1.84 & 3.03 & 5.06 &1.4 \\
 & Halo 2 & 2.26 & 6.56 & 2.25 & 2.91 & 4.13 &1.3\\
 & Halo 3 & 2.60 & 7.03 & 3.38 & 2.10 & 3.51 &1.4\\
\hline
\multirow{3}{*}{No Cutoff} & Halo 1 & 1.94 & 5.78 & 1.94 & 2.97 & 2.97 &1\\ 
 & Halo 2 & 2.93 & 6.63 & 2.22 & 2.98 & 2.98 &1 \\ 
 & Halo 3 & 3.81 & 7.22 & 3.47 & 2.09 & 2.09 &1\\

\hline\hline
\end{tabular}
\end{center}
\end{table}
Dark matter halo concentrations are directly related to the halo formation times, which can be calculated from the mass fluctuation spectrum $\sigma(M)$ (see e.g. \cite{Bullock2001,Kuhlen2005,Maccio2008,Pradaetal2012}). On microhalo scales, $\sigma(M)$ decreases only very slowly with mass, which leads to nearly constant typical concentrations. The microhalo simulations in \cite{Diemand2005} used a box size of 3 comoving kpc which contained a small, very high resolution region where microhalos ($M \sim 10^{-6}M_{\odot}$) were resolved. The surrounding region with lower resolution contained some larger halos ($M \sim M_{\odot}$) which indeed showed similar concentrations as the microhalos. The nearly constant $\sigma(M)$ also means that the scale of the small simulation boxes (or refinement regions) used to resolve microhalos enter the non-linear regime soon after the typical microhalo formation time. This is the reason why microhalo simulations usually end around $z=30$ and usually resolve only a small sample of objects. Finite box size simulations also exclude large scale fluctuations, which are significant in this mass and redshift regime. A 30 pc box for example, as used in this work and in Ishiyama et al. (2010) \cite{Ishiyama2010}, could lower the typical formation redshift by about a factor of two (see Figure 3 in \cite{Ishiyama2010}), which would lower the microhalo concentrations by the same factor. On the other hand, the microhalos which exist in our final snapshot ($z=31$) represent an earlier forming, more concentrated subset of the full microhalo sample.

Besides the limitations mentioned above, we will still use our sample of microhalo concentrations for a rough comparison with CDM mass - concentration relations. At $z = 31$ and $M = 10^{-7}$ M$_{\odot}$, the Bullock et. al (2001) model (\cite{Bullock2001}, in the normalisation of \cite{Kuhlen2005}) predicts a mean concentration $c_{\rm NFW,200} = 1.85$ with a halo-to-halo one sigma scatter of $\Delta(\log c) \simeq 0.18$ at a given mass. A more recent, refined version of the Bullock et al. (2001) model \cite{Maccio2008} predicts $c_{\rm NFW,200} = 2.43$ and the same scatter. Both models are consitent with the $c_{\rm NFW,200}$ values found here (see Table 1) and also with the microhalo samples from earlier microhalo simulations \cite{Diemand2005,Ishiyama2010}.

In the simulations with a cutoff the profiles do not follow NFW and we used steeper functions ($\alpha = 1.3,\; 1.4$). This shifts the best fit scale radius $r_s$ further out and therefore reduces the concentration parameter compared to the NFW case. Keeping the radius $r_{v_{\rm max}}$ of the maximum circular velocity fixed, one can convert the concentrations between an NFW profile and variations of it \cite[e.g.][]{Klypin2001,Ricotti2003}. For the case of $\alpha = 1.3$ and $\alpha = 1.4$ respectively, the conversion is given by
\begin{equation}\label{concentrationrelation}
c_{1.3} = c_{\rm NFW}/1.42, \quad c_{1.4} = c_{\rm NFW}/1.67.
\end{equation}
The $c_{\rm NFW}$ values obtained from this conversion (see Table \ref{halosizetable}) are higher than those of the halos in the no-cutoff model. Apparently the cutoff increases both the microhalo central densities and their concentrations.

After halo formation the typical scale radii of halos of a given mass are roughly constant, while $r_{200}$ and $c_{200}$ increase proportional to the expansion factor $a = 1/(1+z)$ \cite[e.g.][]{Bullock2001}. Assuming this average scaling relation applies also to our individual microhalos, the z=0 concentration estimates for Halo 1-3 in the cutoff model is $c_{\rm 200} = 97.0$, $93.1$ and $67.2$. These are again in line with mean concentrations predicted by the models at z=0: $c_{\rm NFW,200} = 59.2$ (Bullock et al. (2001)) and $c_{\rm NFW,200} = 77.8$ (Macci\`o et al. (2008)).

A simple power law approximation to the Bullock et al. model is often used on galactic halo scales, $c_{\rm NFW,200} \simeq 8.45 \times (M / 10^{12} M_{\odot})^{-0.11}$. Extrapolating this simple approximation all the way down to microhalo scales ($10^{-7}$ M$_{\odot}$) give typical microhalo concentrations of around 1040 at z=0. Even with our small sample of z=31 microhalos (or with any of the earlier microhalo samples from \cite{Diemand2005,Ishiyama2010}) one can clearly rule out such a simple power law concentration-mass relation.  

\section{Small scale structure contributions to the DM annihilation signal}\label{sec:boost}

Dark matter annihilation rates are proportional to the density squared, so any small scale clumpines contributes to indirect dark matter detection signals. The total annihilation luminosity $L$ from a dark matter halo consists of a main halo signal $\tilde{L}$ from the smooth spherically averaged dark matter distribution and from a substructure contribution $L_{\rm sub}$, often expressed as a halo boost factor $B(M)$:
\begin{eqnarray}\label{defboostfactor}
L(M) = \tilde{L}(M) + L_{\rm sub}(M) = \tilde{L}(M) + B(M)\tilde{L}(M) =  \tilde{L}(M) (1 + B(M) ) \nonumber,
\end{eqnarray}

Substructure boost factors play a significant role in the prediction of indirect detection signal, for example when deciding if nearby small dwarf galaxies or  local galaxy clusters are the most promising targets \cite{Sanchez-Conde2011}.  Large $N$-body simulations have shown that the numerically resolved subhalos in galactic halo have $L_{\rm sub} \simeq L_{\rm sub}(M)$, i.e. $B(M)\simeq 1$, and that significant contributions to B(M) from unresolved smaller subhalos gives boost factors of a few up to about ten, depending on details of the extrapolations to microhalo scales \cite{Diemandetal2007,Diemandetal2008}.  On the other hand, simplistic models based on power laws for $c(M)$ or $L_{\rm sub}(>M)$ give much larger boost factors of a few hundred for galactic halos \cite{Springel2008}, see \cite{Kuhlen2012} for a recent review and discussion. The different boost values are mostly caused by very different assumptions about microhalo concentrations. In the following we will show that only the lower estimates (e.g. for galactic halos $B(10^{12} M_{\odot}) \lesssim 20 $) are consistent with simulated microhalo concentrations.

The boost factor of a dark matter halo of mass $M$ is given by the integrated subhalo contribution (form the CDM free streaming cutoff mass $m_0$ up to the main halo mass $M$) divided by the luminosity of the main halo $\tilde{L}$ \citep[e.g.][]{Strigari2007,Pieri2009,Kuhlen2008,Pieri2010,Kuhlen2012,SanchezPrada2013}:
\begin{eqnarray}\label{boostfactor}
B(M) &=& \frac{1}{\tilde{L}(M)} \int_{m_0}^{M} dm \; \frac{dN}{dm} \; L(m)  \\
&=& \frac{A}{\tilde{L}(M) \; M} \int_{m_0}^{M} dm \; \Big(\frac{m}{M}\Big)^{\alpha} \big(1 + B(m)\big) {\tilde{L}(m)} \nonumber,
\end{eqnarray}
where $dN/dm$ is the subhalo mass function and $\alpha$ its slope. The normalisation of the subhalo mass function, $A$, is set by requiring a fraction of the main halo mass to be in subhalos. For $\alpha = -1.9$ it is obtained to be $A\approx 0.03$ \cite[e.g.][]{Zentner2003,Bosch2005}. The total annihilation luminosity coming from a single halo of mass $M$ is given by
\begin{equation}
{\tilde{L}(M)} = 4 \pi M  \frac{c^3}{f(c)^2},
\end{equation}
where $c$ is the concentration parameter and $f(c)$ is fixed by the enclosed mass (and is therefore dependent on the exact form of the density profile) \cite[e.g.][]{BinneyTremaine}. For a two parameter density profile as in Eq. \eqref{densityfit}, the general form of $f(c)$ is given by
\begin{equation}\label{fofc}
f(c) = \Big(\frac{c^{3-\alpha}}{3-\alpha}\Big)\times {}_2 F_1(3-\alpha;\; 3-\alpha;\; 4-\alpha;\; -c),
\end{equation}
where ${}_2 F_1$ is the hypergeometric function. In the case of a NFW profile ($\alpha=1$, $\gamma=3$), Eq. \eqref{fofc} reduces to $f(c)= \log(1+c) - c/(1+c)$. To evaluate $B(M)$ a one needs to know the form of the subhalo density profiles and also thier concentrations.
 
\subsection{Moderate microhalo concentrations rule out very large boosts from power-law extrapolations}

For the subhalo mass function the extrapolation of a simple power law is consistent with theoretical models (e.g. \cite{Giocoli2009} and with microhalo simulations \cite{Diemand2005,Diemand2006}). The same is not the case for the concentration--mass relation $c(M)$: models and microhalo simulations are consistent with concentrations related to halo formation times and therefore to the mass fluctuation spectrum $\sigma(M)$ and to the power spectrum $P(k)$, which are not power laws (see section \ref{concentrations}). In realistic models one combines a power law subhalo abundance with a non-power law concentration model and gets a total subhalo luminosity $L_{\rm sub}(m_0)$, which is not a simple power law of the minimum subhalo mass $m_0$. The very large boost factors (over 200 for galaxy halos \cite{Springel2008} and over 700 for clusters \cite{Gaoetal2012}) obtained from power law extrapolations are ruled out by these and earlier microhalo simulations \cite{Diemand2005,Ishiyama2010} and by the typical halo formation times expected from the CDM mass fluctuation spectrum $\sigma(M)$.

\begin{figure}
\begin{center}
\includegraphics[scale=0.33]{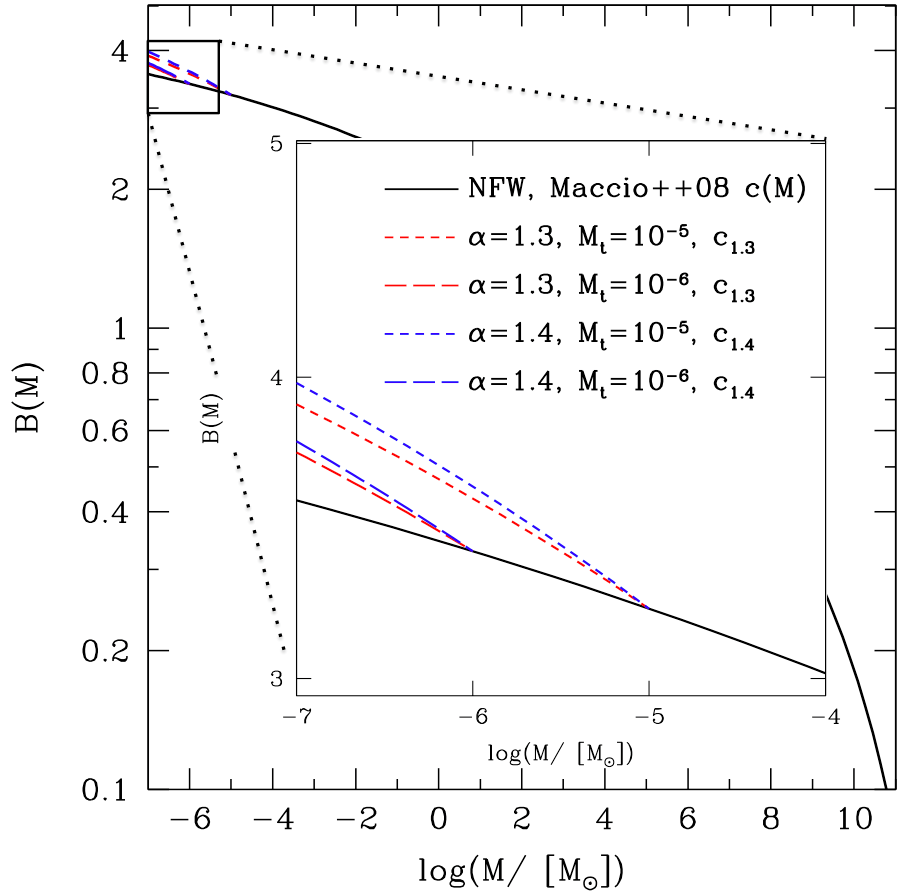}
\caption{The annihilation luminosity boost factor extrapolated down to a cutoff mass of $m_0=10^{-7}$M$_{\odot}$. The black solid line shows the boost factor results under the assumption of NFW density profile ($\alpha=1$) in combination with a concentration-mass relation as suggested by \cite{Maccio2008}. The dashed colored lines show the additional contribution to the boost factor when the inner density profile of the smallest microhalos has slope of $\alpha=1.3 \;({\rm red}),\; 1.4 \;({\rm blue})$ and transition scales of $10^{-5}$M$_{\odot}$ (short-dashed) and $10^{-6}$M$_{\odot}$ (long-dashed) respectively.}
\label{boostfactors}
\end{center}
\end{figure}

We use the concentration - mass relation of Macci\`o et al. (2008) \cite{Maccio2008}, which is consistent with microhalo simulation results (see section \ref{concentrations}). Assuming universal NFW profiles for all subhalos down to $m_0 = 10^{-7}$ M$_{\odot}$ (and up to $M = 10^{12}$ M$_{\odot}$) gives the moderate value of $B(M) = 3.56$, as indicated by the black solid line in Figure \ref{boostfactors} (which is in agreement with the results of other authors \cite[e.g.][]{Diemand2005,Strigari2007,Kuhlen2008}). This will be our reference value derived from a simple, plausible subhalo model. Slightly different models are possible given our current understanding of structure formation and could include the following variations which change the boosts by small factors:

\begin{itemize}
\item Instead of NFW one could use the Einasto profile (with a fixed $\alpha_E = 0.17$) instead. It fits density profiles of large CDM halos better than NFW \cite[e.g.][]{Navarro2004,Merritt2006}, it is denser than NFW in the inner halo (around 0.003 to 0.1 $r_s$) and normalised to the same outer profile and concentration gives about 1.4 times larger annihilation fluxes\cite{Kuhlen2010,DiemandMoore2011}. 

\item We neglect halo-to-halo scatter in the concentration parameter \cite{Bullock2001,Maccio2008} and the profile shape \cite{Navarro2004,Diemand2004profiles} (i.e. in the second parameter $\alpha_E$, when the Einasto form is used), which could lead to a small increase in the boost factors.

\item The subhalo mass functions might be slightly steeper, even the critical value of $\alpha = -2.0$ (i.e. constant mass in subhalos per mass decade) is consistent with some simulated CDM halos \cite{Diemandetal2007,Gaoetal2012}. Changing $\alpha$ from -1.9 to -2.0 in our reference NFW based model would increase the boost by about a factor of 6 ($B(M)=21.7$).

\item We neglect caustics in microhalos. Their contribution to the $z=0$ annihilation signal is uncertain: on the one hand the effect our simulations show at $z=31$ might increase with numerical resolution and on the other hand the evolution from $z=31$ until present time is unclear. However, since their moderate contribution at $z=31$ appears to decay quickly with time (see Figure \ref{densitymaps} and Section \ref{nbodysims}), we do not expect them to affect $z=0$ boost factors significantly.

\item Steeper inner density profiles for subhalos near the cutoff scale, as found in Ishiyama et al. (2010) \cite{Ishiyama2010} and in the simulations presented here, increase the total annihilation boost factor moderately, as described and quantified in the next section.
\end{itemize}

\subsection{Steeper microhalo inner density profiles increase boost factors moderately}

Microhalos near the cutoff scale have steep central density cusps with a slope of $\alpha = 1.3 - 1.4$. In order to quantify the contribution of such objects to the total annihilation boost, we solve Eq. \eqref{boostfactor} numerically with the boundary condition $B(m_0)=0$. In addition, we introduce a transition mass $M_t = \{10^{-6}, 10^{-5}\}$ M$_{\odot}$ as a sharp transition scale: above $M_t$ the density profiles are assumed to be NFW like, below $M_t$ they are assumed to have the functional form found in Section \ref{sec:profiles}. Fixing the cutoff scale to be $m_0=10^{-7}$ M$_{\odot}$, the results are shown in  Figure \ref{boostfactors}. The dashed colored lines show the increase of $B(M)$ when the smallest objects possess a slope of the inner density profile of $\alpha = 1.3$ (red) and  $\alpha = 1.4$ (blue), with transition scales of $M_t = 10^{-5}$M$_{\odot}$ (short-dashed) and $M_t = 10^{-6}$ M$_{\odot}$ (long-dashed). 
We have incorporated the concentration correction (see Eq. \eqref{concentrationrelation}) in the calculations of $B(M)$. It is evident from Figure \ref{boostfactors} that such steep microhalo cusps increase the total annihiliation boost for a Milky Way sized halo between 5 to 12 per cent, strongly dependent on the inner slope, the transition mass and the cutoff scale.

In the analysis above, we have assumed that the slope of the inner density profile stays unperturbed between $z=31$ and $z=0$. Unfortunately, current state of the art simulations do not allow to evolve a microhalo in a realistic environment to the present epoch. However, there is good reason to assume that the inner density profile of microhalos stays unchanged until $z=0$: on the one hand, it was shown in earlier works, that the inner density profile of larger virialised halos is practically unaffected by mass accretion \cite[e.g.][]{Tasitsiomietal2004,DiemandKuhlenMadau2007}. Furthermore recent results show that in mergers of collisionless dark matter halos the cusp of the steeper progenitor is preserved \cite[e.g.][]{Dehnen2005,Zempetal2008}.

\section{Summary \& Conclusion}
We have used ultra high resolution $N$-body simulations to examine the slope of the inner density profile of microhalos and the associated implications for the total annihilation luminosity boost factor of a galaxy sized dark matter halo. Our results can be summarised as follows:
\begin{itemize}

  \item When introducing an exponential cutoff in the matter power spectrum at a scale that corresponds to the size of a typical microhalo, we found that the density profile of these objects is accurately described by a NFW-like fitting function (see Eq. \eqref{densityfit}) with an inner slope of $\alpha = 1.3 - 1.4$. Without the cutoff, the inner structure is found to be very similar to the one of larger, i.e. galaxy- and cluster-sized, dark matter halos, and are reasonably well fitted by the usual NFW form.
  
  \item Assuming that one (or two) order of magnitude above the cutoff scale the inner slope of the density profile rapidly changes from $\alpha = 1$ to $\alpha = 1.4$, we find that the total annihilation boost factor of a Milky Way sized halo is increased by 5.8 (11.9) per cent. If the inner slope is $\alpha = 1.3$, the increase is found to be 4.7 (9.6) per cent.
  
  \item The measured concentrations of microhalos agree with the concentration - mass relation of Macci\`o et al. (2008) \cite{Maccio2008} (and therefore also with the Bullock et al. (2001) model \cite{Bullock2001}). Power law models for the concentration - mass relation, which would boost the total annihilation luminosity by two to three orders of magnitude, are therefore clearly ruled out.
  
\end{itemize}
Larger microhalos simulations, like the recently presented trillion particle simulation \cite{Ishiyama2012}, will be needed to reduce some remaining uncertainties, especially about the transition mass scale, at which the inner slope of the density profile significantly increases, the scatter in microhalo profiles and the exact mean microhalo concentrations and scatter. More work is also required to understand the relations between inner profiles, halo formation histories and cutoff scales.

\acknowledgments  
We would like to thank Mike Kuhlen, Miguel S\'anchez-Conde and Neal Dalal for helpful discussions and comments in the final stages of this work. Numerical simulations were performed on the SuperMUC at the LRZ in Garching, on the zBox3 at the Institute for Theoretical Physics in Zurich and on the Rosa cluster at the CSCS in Lugano.This work was supported by the SNF.

\end{document}